\documentclass[10pt,conference]{IEEEtran}
\pdfoutput=1
\usepackage{epsfig}
\usepackage[usenames]{color}
\usepackage{amsmath}
\usepackage{amssymb}
\usepackage{mathrsfs}
\usepackage{amsfonts}
\usepackage{lscape}
\usepackage{graphicx}
\newcommand{\Capacity}{\mathcal{C}}

\newcommand{\acal}{a}%{\hbox{\aa}}

\newcommand{\Hcal}{\mathcal{H}}

\newcommand{\Zcal}{\mathcal{Z}}

\newcommand{\tL}{\tilde \Lambda}

\newcommand{\+}{\!+\!}

\newcommand{{\main}}{\hbox{\textsl{main}}}

\newcommand{\intersect}{\cap}
\newcommand{\union}{\cup}

\begin{document}

\title{Toward Fast Reliable Communication at Rates Near Capacity with Gaussian Noise
\vspace{-.3cm}
}

% use a multiple column layout for up to three different affiliations

\author{\authorblockN{Andrew R Barron, Antony Joseph}
\authorblockA{Department of Statistics, Yale University
and SumCodes, a Division of Barron Field, LLC\\
Email: {andrew.barron@yale.edu, antony.joseph@yale.edu}\\
Presented at the {\em IEEE International Symposium on Information Theory}, June 13-18, 2010.
\vspace{-.5cm}
}}
% avoiding spaces at the end of the author lines is not a problem with
% conference papers because we don't use \thanks or \IEEEmembership

\maketitle

\begin{abstract}
For the additive Gaussian noise channel with average codeword power constraint, sparse superposition codes and adaptive successive decoding is developed. Codewords are linear combinations of subsets of vectors, with the message indexed by the choice of subset. A feasible decoding algorithm is presented. Communication is reliable with error probability exponentially small for all rates below the Shannon capacity.
%This provides mathematically proven practical solution to Shannon's Gaussian channel communication problem.
\end{abstract}

\vspace{-.1cm}
\section{Introduction}
\label{sec:intro}
Sparse superposition codes with computationally feasible decoding is shown to achieve exponentially small error probability for any %communication
rate below the %Shannon
capacity.  A companion presentation %\cite{BarronJosephLeastSquares}
\cite{BarronJosephLeastSquaresISIT}
gives bounds for optimal least squares decoding. %Both merge modern perspectives on statistical regression and information theory.

Code construction is by linear combination of vectors of length $n$.  Let $X_1,X_2,\ldots,X_N$ be a dictionary of such vectors. %, each of length $n$.
Organize it in a matrix $X$ of $N\!=\!BL$ columns, partitioned into $L$ sections of size $B$ a power of $2$. Codewords are superpositions $X\beta=\sum_j \beta_j X_j$ with each section having $1$ term %coefficient
non-zero. %, with magnitude to be specified.
The set of such $\beta$ is not closed under linear combination, so these are not linear codes in the algebraic coding sense.  Nevertheless, they are fast to code and decode.
%, with known value of $\beta_j$ if it be non-zero.

The message is conveyed by the choice of the subset of $L$ terms, with one from each section.  From an input bit string $u=(u_1,u_2,\ldots,u_K)$, with $K=L\log_2 B$, encoding is realized by %interpreting
regarding $u$ as a concatenation of $L$ numbers, each with $\log B$ bits, specifying the selected columns.  The codewords $c=X\beta$ have power $(1/n)\sum_{i=1}^n c_i^2$, which will be near $P$ when averaged across the $2^K$ possible codewords. The received vector is $Y=X\beta + \epsilon$ with $\epsilon$ distributed N($0,\sigma^2 I$).

A decoder maps the received vector into an estimate $\hat u$.  With $sent=(j_1,j_2,\ldots,j_L)$ being the terms sent, the decoder produces estimates $\hat j_1,\hat j_2,\ldots,\hat j_L$. Overall block error is the event $\hat u \neq u$ and section error is the event $\hat j_\ell\ne j_\ell$. The fraction of section mistakes is $(1/L)\sum_{\ell=1}^L 1_{\{\hat j_\ell \neq j_\ell\}}$.

The reliability requirement is that the mistake rate is small with high probability or the block error probability is small, averaged over input strings $u$ as well as the distribution of $Y$. The supremum of reliable communication rates $R\!=\!K/n$ is the channel capacity $\Capacity \!=\! (1/2) \log_2 (1 \!+\! P/\sigma^2)$, as in \cite{Shannon}, \cite{CoverThomas}.

The challenge is to achieve arbitrary rates below the capacity, with reliable decoding in manageable computation time.
%The present paper takes steps toward meeting this challenge.
Here communication rates are identified which are moderately close to the capacity and a fast decoding scheme is devised. It is demonstrated to have probability that is exponentially small in $L/(\log B)^2$ of there being more than a moderately small fraction of section mistakes.

The setting adopted is
%The development here is for
the discrete-time channel with real-valued inputs and outputs and independent Gaussian noise.  Standard communication models have been reduced to this setting as in \cite{Gallager1968}, \cite{ForneyUngerboeck}, when there is a frequency band constraint with specified noise spectrum.
%so
Solution to the coding problem, married to appropriate modulation, is
%regarded as
relevant to myriad settings involving transmission over wires or cables for internet, television, or telephone or in wireless radio, TV, phone, satellite or other space communications.  Previous standard approaches, as discussed in \cite{ForneyUngerboeck}, entail a decomposition %of the problem
into separate problems of modulation, of shaping of a multivariate signal constellation, and of coding.  %As they point out,
Though there are practical schemes with empirically good performance, theory for practical schemes achieving capacity is lacking.  In our analysis, shaping is built directly into the code design.

%Practical coding has also been concerned with channels in which either the code symbols take on only two possible values, such as the binary symmetric channel. These arise from a continuous-valued channel by constraining $c_i$ to the values $\pm \sqrt P$.  With such binary signaling, the available capacity is not more than $1$. When the signal-to-noise ratio is not low, rather than making such restrictions, the preference is to allow higher rates of communication, by solving the less constrained Gaussian channel coding problem.

The entries of $X$ are generated with the independent standard normal distribution.  The coefficients are $\beta_j$ equal to $\sqrt{P_{(\ell)}}$ for $j=j_\ell$ in $sent$ and equal to $0$ otherwise, with sum of squares $\sum_{\ell=1}^L P_{(\ell)} = P$ matching the power constraint. In the simplest case, the same power is allocated to each section $P_{(\ell)} = P/L$.  We also consider the choice of variable power with $P_{(\ell)}$ proportional to $e^{-2C\ell/L}$ and a slight variant of this allocation in which the power is variable across most $\ell$ and then levels for $\ell/L$ near $1$.

For a rate $R$ code, $nR = L\log B$, so the codelength $n$ and the subset size $L$ agree to within a log factor.  Setting $L=B$ is sensible, or, for a target codelength $n$, one may set $B=n$ and $L=nR/\log n$.  For the best case developed here, the rate $R$ is chosen to have a drop from capacity that is near $1/\log B$, to within a loglog factor. %as a function of $B$, such that the gap from capacity is small.
When the signal to noise ratio is large, one finds it desirable to arrange $\log B$ to be at least as large as $C$ to achieve at least a constant fraction of capacity.

%Possible dictionary sizes are between $K$ and $2^K$. In one extreme, with $1$ section of size $B\!=\!2^K$, the dictionary $X$ is the whole codebook.  At the other extreme there would be $L\!=\!K$ sections, each with two candidate terms in subset coding or two signs of a single term; in which case $X$ is the generator matrix of a linear code. Between these extremes, we construct feasible, reliable, high-rate codes with codewords corresponding to linear combinations of subsets of a moderate number of terms.

%The least squares problem of seeking $\beta$ to minimize $\|Y  -  X\beta\|^2$ subject to convex constraints is made more challenging by the non-convex optimization requirement of a specified number of non-zero terms, one in each section.

%Convex projection, and its associated algorithms, finds $X\beta$ in the convex hull that comes closest to $Y$, from which one can move to a nearby vertex.  Convex projection has a rate limitation that does not allow it to directly achieve capacity.  Nevertheless, certain iterations associated with such algorithms are motivational in defining revised algorithms that operate on the vertices.

Let's summarize our findings. With constant power allocation, %when the power of the terms sent are the same at $P/L$,
a two-step algorithm and a multi-step improvement reliably achieve rates up to a rate $R_{0} = (1/2)P/(P\!+\!\sigma^2)$ less than capacity.
%The number of steps required is of constant order.
With variable power and order $\log B$ steps, we bring the achievable rate up near capacity ${\Capacity}$, albeit with a gap from capacity of order $1/\sqrt{\log B}$.  With the variant in which the power is leveled for $\ell/L$ near $1$, the gap from capacity is reduced to order $1/\log B$, to within a loglog factor, and, moreover, the section mistake rate is less than a constant times $1/\log B$, except in an event of probability exponentially small in $L/(\log B)^2$, as we report here.  Subsequent to the submission of this conference paper, we have refined 
%the techniques to improve this error 
this probability bound, obtaining that it is exponentially small in $L/(\log B)$, or equivalently $n/(\log n)^2$, to within a loglog factor, as will be reported in the upcoming journal submission.
%Our tools provide reasonable constants, though the best values of these for our scheme are not known.

The performance, as measured by the gap from capacity at a similar reliability level, is comparable to benchmarks of performance for schemes not demonstrated to be practical, including \cite{BarronJosephLeastSquaresISIT} for least squares decoding of related superposition codes, and \cite{PolyanskiyPoorVerdu} for theoretically optimal codes. For a gap from capacity of order $1/\log n$, the best error probability is exponentially small in $n/(\log n)^2$.
%, though it is not known whether than probability is achievable by practical schemes.
%There is potentially room for improvement in our results in terms of constants and avoidance of the loglog factor.
%As our probability is exponentially small in $L/(\log n)^2$, which is $n/(\log n)^3$, there is potentially some room for improvement.

The decoder initially computes for the received $Y$, its inner product with %each of
the terms in the dictionary, and sees which are above a threshold. Such a set of inner products and comparisons is performed in parallel by a basic computational unit, e.g. a signal-processing chip with $N$ parallel accumulators, in time of order $n$. These are pipelined so that the inner products are updated in constant time as each element of $Y$ arrives.

The threshold, set high enough that incorrect terms are unlikely to be above threshold, leads to only a small fraction of terms decoded in any one such step.  Additional steps are used to bring the total fraction decoded %up
near $1$. These steps take the inner products with residuals of the fit from the terms previously above threshold. A variant of the %idea of
inner product with residuals is found to be somewhat more amenable to analysis.

The decoder does not predetermine which sections %of terms
are to be decoded on any one step, rather it adapts the choice in accordance with which has inner product observed to be above threshold.  Thus we call it
%our class of procedure
{\em{adaptive successive decoding}}.
%A variable power allocation is studied with $P_{(\ell)}$ proportional to $e^{-2\Capacity\ell/L}$.  It is motivated by rate splitting considerations, but for our reliability properties it is critical that we are decoding sections simultaneously. For the moderate dictionary sizes one can not reliably solve for the terms one section at a time.

We determine a function $g(x)$ mapping from $[0,1]$ into $[0,1]$, which has the role that if $x_{k-1}$ is a likely fraction of sections correctly decoded from previous steps up to $k\!-\!1$ then $g(x_{k-1})$, slightly adjusted, provides a value $x_k$ of total fraction of sections likely to be correctly decoded by step $k$. This function depends on the power allocation rule and the choice of rate. A choice of communication rate is acceptable if the function $g(x)$ is greater than $x$ over most of the interval.  Such a function $g$ is said to be \emph{accumulative}, allowing the succession of steps to build up a large fraction of correctly decoded sections, with only a small fraction of mistakes remaining.  The role of $g(x)$ is illustrated in Figure $1$.
%for correct decoding of terms to continue until most are correctly decoded.

Our analysis provides summary formulas for the rate and the target fraction of mistakes %, for particular power allocation rules,
that arise from bounding the extent of positivity of $g(x)-x$. These summary formulas provide proof of a favorable scaling of rate by our scheme for the particular reliability targets, indexed by the size of the code.

\begin{figure}
\vspace{-.3cm}
\centerline{
\mbox{
\includegraphics[width=1.9in]{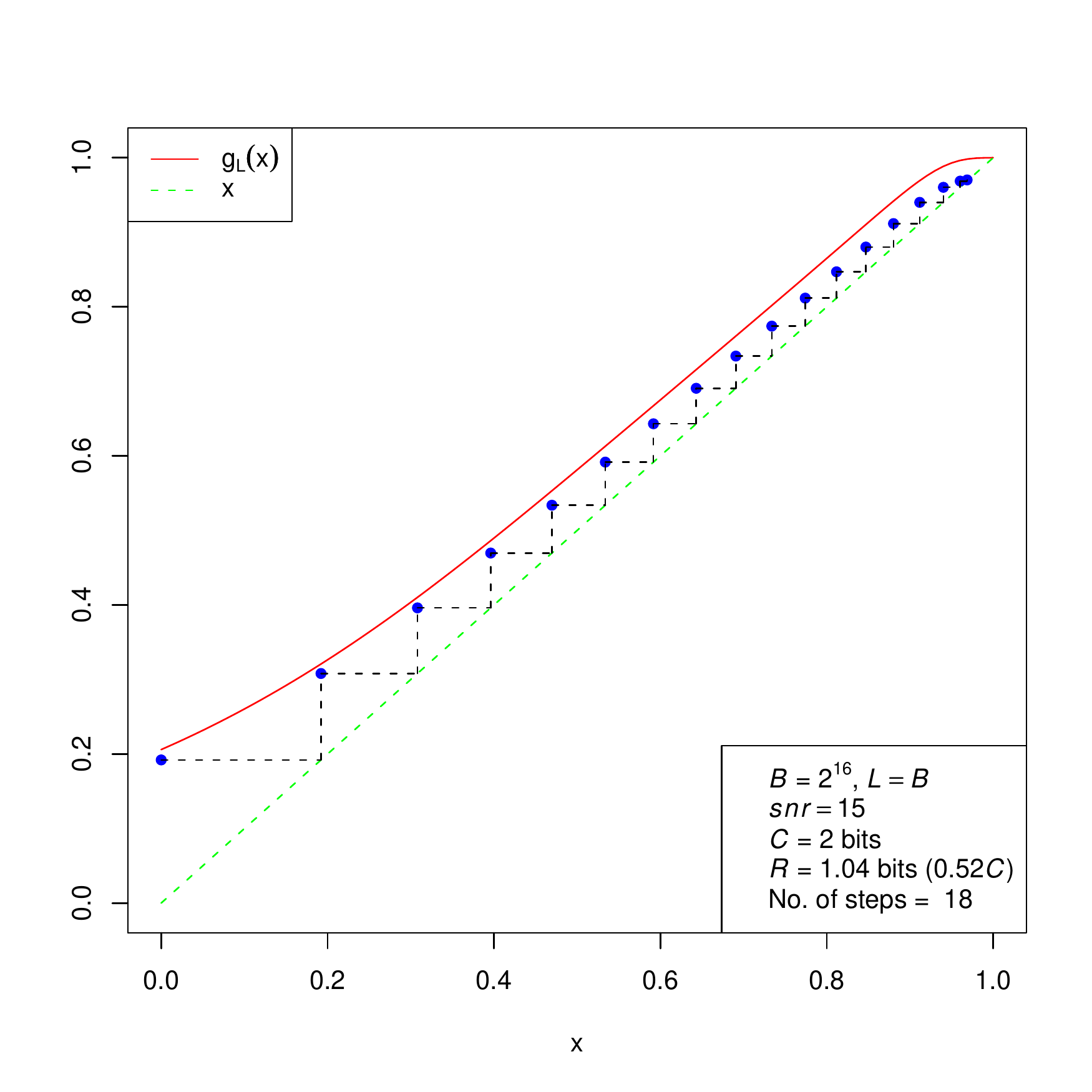}
}
\!\!\!\!\!\!\!\mbox{
\includegraphics[width=1.9in]{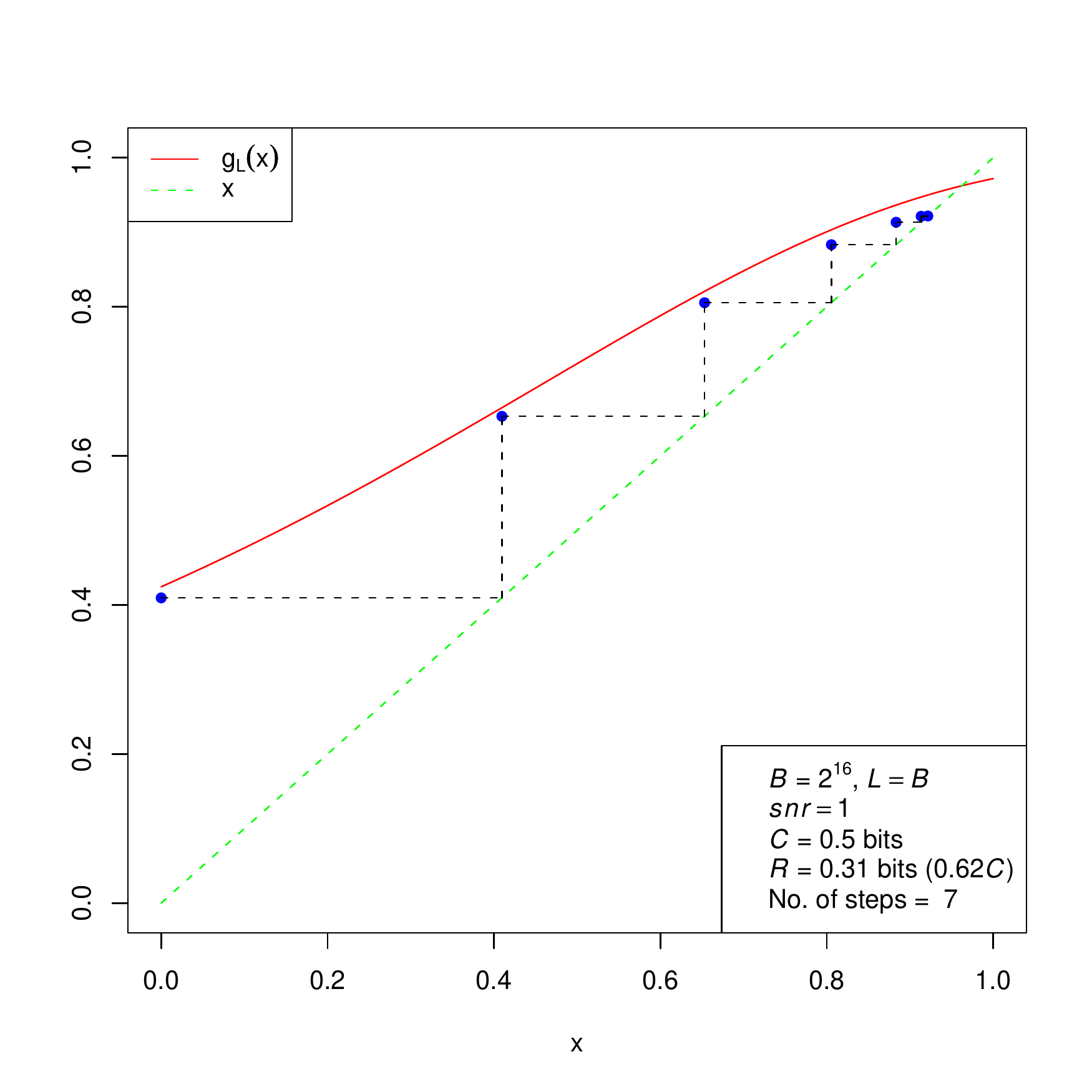}
}
\vspace{-.5cm}
}
\caption{
%\vspace{-.5cm}
Plots of $g(x)$ and the sequence $x_k$. For $snr = 15$ %The dots indicate the sequence $q_{1,k}^{adj}$ for the
%18 steps. %Here $L = 6.5536\times 10^4$, $B = 2^{16},\, snr = 15$
%and $R = 1.04$.
the plot takes $a =0.86$, $c = 1.6$ and $\gamma = 0.8C$
%Weighted (un-weighted) detection rate is $0.995$ ($0.987$)
and the final
false alarm and failed detection rates are $0.026$ and  $0.013$ respectively, with probability bound of at least that fraction of mistakes equal to $0.002$.
%The above occurs except in a set of probability less than  $2.1\times 10^{-3}$.
For $snr=1$, constant power allocation is used with $a =0.56$
%The detection rate (both weighted and un-weighted) is $0.947$
and the
false alarm and failed detection rates are $0.026$ and  $0.053$ respectively, with probability bound $0.0007$.
\vspace{-.5cm}
}
\label{fig:prog15}
\end{figure}

Moreover, the function $g(x)$ can be evaluated in detail to choose settings of parameters ($a$, $c$, and $\gamma$ below). This allows computation of the best communication rate our analysis achieves, for given error probability and target mistake rates.

%There are two key non-negative parameters that arise in our scheme, denoted $a$ and $c$.
The parameter $a$ arises in %setting
the threshold $\tau=\sqrt{2\log B}+a$ of the standardized inner products.  The parameter $c$ sets the height at which the variable power is leveled, with power $P_{(\ell)}$ chosen to be proportional to $\max\{e^{-2\Capacity(\ell-1)/L}, cut\}$, with $cut= e^{-2\Capacity} (1\+\delta_c)$ where $\delta_c=c /\sqrt{2\log B}$.

%For additional flexibility
Allowing power proportional to
$\max\{e^{-2\gamma(\ell-1)/L}, cut\}$, with $cut= e^{-2\gamma} (1\+\delta_c)$, for $\gamma$ between $0$ and $\Capacity$, %yields additional flexibility,
interpolates between the constant and variable cases.
%allocation that arises when $\gamma=0$ and the variable case with $\gamma=\Capacity$.

Figure $2$ plots the rate $R$ as a function of $B$, from optimization of $a$, $c$, and $\gamma$, maintaining the bound $10^{-3}$ on the probability of a fraction of mistakes exceed $0.10$.  Both the case $L=B$, and a large $L$ limit are shown as well as some results of simulation of the algorithm with $L=100$.

Signed superposition coding in which the $\ell$'th non-zero coefficient value is $\pm \sqrt{P_{(\ell)}}$ increases the number of codewords to $(2B)^L$
%and to near $(eB)^L$, respectively,
with the same reliability bounds, thereby improving the rate by a factor of $1+(\log 2)/(\log B)$, above what is shown in Figure 2.  %Further improvement using
Arbitrary $L$ term subset coding (without partitioning) is possible, though not as simple, for a total rate improvement by a $1+(\log 2e)/(\log B)$ factor.
For this presentation, we focus on the unsigned, partitioned superposition code case.

%Signed superposition coding in which the $\ell$'th non-zero coefficient value is $\pm \sqrt{P_{(\ell)}}$ and arbitrary $L$ term subset coding (without partitioning) may be used to increase the number of codewords to near $(2eB)^L$
%and to near $(eB)^L$, respectively,
%with the same reliability bounds.
%There is potential for further increase with arbitrary $L$ term subset coding (without partitioning)
%These modifications improve the rate by a factor of $1+(\log 2e)/(\log B)$, above what is shown in Figure 2.

\begin{figure}[t]
\vspace{-0.2cm}
\centerline{
\mbox{
\includegraphics[width=1.7in]{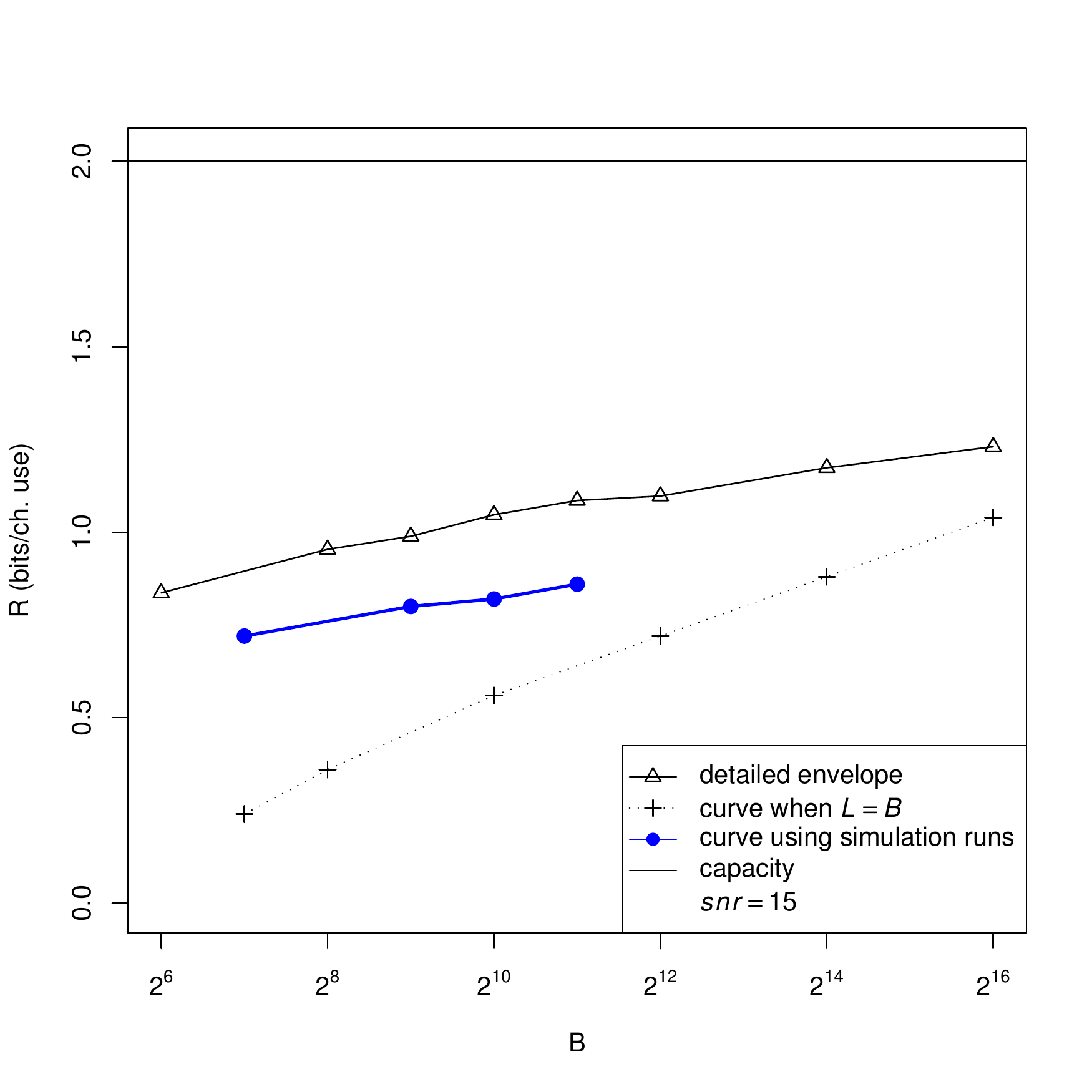}
}
\mbox{
\includegraphics[width=1.7in]{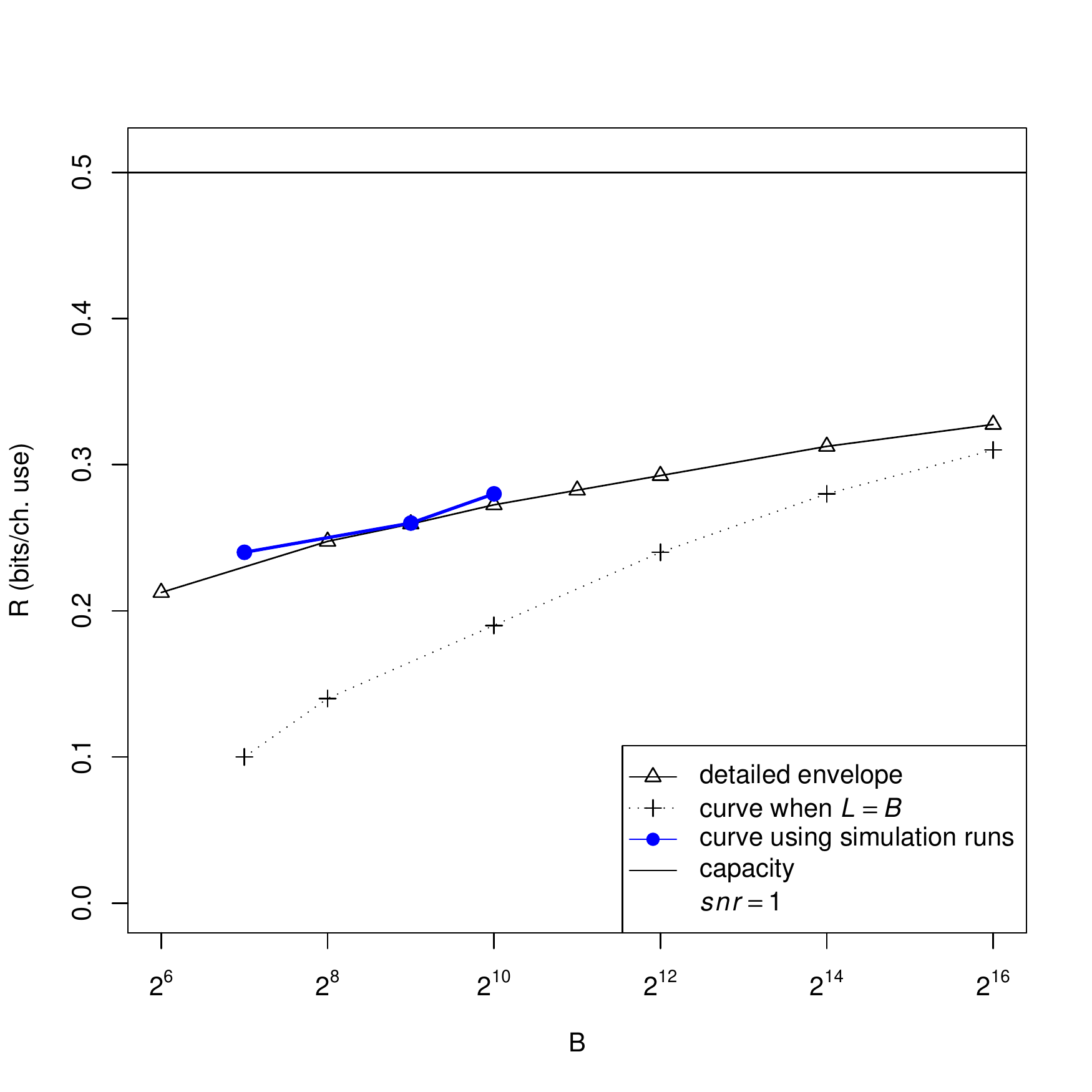}
}
\vspace{-0.4cm}
}
\caption{
%\vspace{-0.4cm}
Curve showing achieved rates as a function of $B$ for $snr = 15$ and $snr = 1$. The x-axis has $B$ plotted on the $\log$ scale.
\vspace{-0.4cm}
}
\label{fig:aymp15}
\end{figure}

%The analysis suggests a role for the rate $R_\gamma^* = \gamma (1-e^{-2\Capacity})/(1-e^{-2\gamma})$, which likewise interpolates between $R_0$ and the capacity $\Capacity$.  Indeed, specifying the rate $R_\gamma$ to be a fraction (preferably near $1$) of the $R_\gamma^*$, one finds that the resulting for instance $R_\gamma= R_\gamma^*/(1+a/\sqrt{2\log B})^2$.

%Our main conclusion is that with rate $R$ near the capacity ${\Capacity}$, after order $\log B$ steps, the fraction of mistakes is small, less than order $e^{-a \sqrt{2 \log B}}$, except in an event of probability nearly exponentially small in $L$, with $a$ equal to a positive constant free to be specified. The fraction of mistakes is smaller than any polynomial in $1/\log B$, and hence smaller than a polynomial in $1/\log n$, sufficient for subsequent correction of these mistakes by an outer code.  The best rate we achieve is $$R = \frac{C}{1+ 4\, \rho \, \epsilon_B}$$ where $\rho = (1\!-\!e^{-\Capacity})/(1\!-\!e^{-2\Capacity})$ is between $1/2$ and $1$ and where $\epsilon_B = a/\sqrt{2\log B}$.  For instance, with $a=1$ and $B=2^{18}=256$kilobits, near $e^{12.5}$, the rate is between $55\%$ and $71\%$ of capacity.  This $R$ represents what we call a practical rate below capacity.  In some cases it is not as close to $\Capacity$ as for the theoretically optimum superposition decoding with similar sizes of $B$, for which bounds are available from our companion paper.

%There is reliability advantage to using smaller $R \!<\! {\Capacity}$, as then the fraction of mistakes becomes of the order of a polynomial in $1/B$.

To prevent block errors, our subset superposition codes combine with error correction codes. The idea is to arrange sufficient distance between the subsets. Consider composition with an outer Reed-Solomon (RS) code of rate $1-2\delta$ near one, for an overall rate $(1\!-\!2\delta)R$.  The alphabet of the RS code %symbols
is taken to be of size $B$. Interpret its codewords %are interpreted
as providing the sequence of labels $j_1,j_2,\ldots,j_L$ of the terms selected from the sections. The RS codelength $L$ is taken to be either $B-1$ or $B$ using a standard
%single symbol
extension.  RS code properties as in \cite{MacWilliamsSloane1977} guarantee correction of any fraction of section mistakes less than $\delta$.
%A forerunner providing
For advocacy of code concatenation %the use of such outer codes is the work of
see \cite{Forney1966}.
%The outer code corrects the few remaining mistakes so that we end up not only with small bit error rate but also with small block error probability.
%Operating on sequences of batches of bits, the fraction of mistakes guaranteed corrected is of order $\delta/\log n$, as shown in \cite{MacWilliamsSloane1977}.  Though not a great correction rate, combined with the superposition code it is certainly enough for our purposes.  Match $\delta/\log n$ to $e^{-a\sqrt{\log B}}$ and the set inner code rate $R$ near ${\Capacity}$ as above.
As a consequence of our result for the inner code, the composite code makes no mistakes, except in an event inheriting the exponentially small probability in $L/(\log B)^2$.
%Moreover, the overall code rate $(1\!-\!\delta)R$ remains close to capacity.  The section size $B$ required for $e^{-a\sqrt{\log B}}$ to match $\delta/\log n$ is $B$ of the order of an exponential of $(\log(1/\delta))^2$.  This is not as good as a polynomial in $1/\delta$, yet considerably better than an exponential in $1/\delta$.

%The analysis of concatenated codes in \cite{Forney1966} is an important forerunner to the development we give here.

A fascinating alternative approach is channel polarization \cite{Arikan2009,ArikanTelatar2008}, which achieves high rates for binary signaling with feasible decoding, with error probability exponentially small in $n^{1/2}$.  For our scheme the error probability is exponentially small in $n^{1-\epsilon}$ for any $\epsilon \!>\! 0$ and communication is permitted at higher rates beyond that associated with binary signalling.

%Concerning other work, there has been progress for binary input channels.
Codes empirically demonstrated to be good include low density parity check codes and turbo codes, both with iterative statistical belief propagation decoding, but mathematically proof of performance near capacity is so far limited to special cases such as the binary erasure channel
%, for low density parity check codes, is empirically shown to provide reliable and moderately fast decoding at rates near the capacity for such channels, and mathematically proven to in certain special cases, such as the binary erasure channel
\cite{LMSS2001a,LMSS2001b}.
%Belief networks are also used for empirically good turbo codes that allow real-valued received symbols, with a discrete set of input levels.  Though steps have been made, as far as we are aware there is not mathematical proof of the desired properties at rates near capacity for the Gaussian channel.

Another approach to sparse superposition decoding is convex projection with  $\ell_1$ constraint, arising from analogous problems of statistical learning and signal recovery. Iterative procedures and properties for such projection are in \cite{Jones1992},\cite{Barron1993},\cite{LBW1996},\cite{BCDD2007}, \cite{HBC}, with preliminary findings for communication in \cite{Tropp2006}.
%., with bounds on the accuracy of the fit to the projection.  Each pass of those greedy algorithm is analogous to the steps of our algorithm.
%Adapted to the present setting,
Each iteration would find in each section the term of highest inner product with the residuals and use it to update the convex combination. %Its computation is linear in the product of the size of the dictionary and the number iterations.
%Projection analysis in the equal power case shows that an approximation to the projection has largest weight in most sections at the term sent, provided the rate $R$ is less than $R_{0}$.
It is unclear to us whether convex projection for communication can be reliable at rates up to capacity. %with the variable power allocation.
%; whereas for larger rates the weights of the projection are too spread across the terms in the sections to identify what was sent.

The conclusions may be expressed in the language of sparse signal recovery.  $L$ terms from a dictionary are linearly combined and subject to noise. For signals $X\beta$ recovery of the terms from the received noisy $Y$ is possible provided the number of observations $n$ is at least  $(1/R) L \log B$.  Recovery using $\ell_1$ constrained convex optimization is accurate provided $R \!<\! R_{0}$ in the equal power case. For our variable power designs, our results establish recovery by other means at higher $R \!<\! {\Capacity}$.
These findings complement work in \cite{Wainwright2009a},\cite{Wainwright2009b},\cite{FletcherRanganGoyal2009},\cite{DonohoEladTemlyakov2006},\cite{CandesPalm2009},
\cite{Rad},\cite{KHMV}. For typical signal recovery problems there is greater freedom of design with non-zero coefficients values regarded as unknown, leading to bounds based on the minimum non-zero signal size, rather than exclusively based on the total signal power as in %identification of
the communication capacity.
%that translates into achieving positive rate, but not capacity.

Superposition codes began with \cite{Cover1972} for the broadcast channel, and later for multiple-access channels \cite{CaoYeh2007},\cite{RimoldiUrbanke2001}.
Our purpose of computational feasibility is different from the original purpose of identifying the set of achievable rates.  Another connection is the consideration of rate splitting and successive decoding. Our adaptive successive decoding yields feasibility in the single-user case and should work also in multi-user settings.
%as well.
%However, feasibility has been lacking in rate splitting analysis in the absence of demonstration of reliability at high rate with superpositions from polynomial size dictionaries with moderate computational complexity.  As said, adaptation rather than pre-specification of the set of sections decoded each step is key to the reliability and speed of our scheme.

%For vectors $a,b$ of length $n$, let $\|a\|^2$ be the sum of squares of coordinates, let $|a|^2 = (1/n)\sum_{i=1}^n a_i^2$ be the average square and let respectively $a^Tb$ and $a \cdot b = (1/n)\sum_{i=1}^n a_i b_i$ be the associated inner products.   Concerning the base of the logarithm ($\log$) and associated exponential ($\exp$), base $2$ is most suitable for interpretation and base $e$ most suitable for the calculus.  Base $e$ is used for convenience in the derivations.

%The dictionary is randomly generated as we said.
%For the purpose of analysis of average probability of at least certain numbers of mistakes,
%We investigate properties with respect to the joint distribution of the dictionary and the noise.

%\clearpage
\section{The Decoder}

From the received $Y$ and knowledge of the dictionary, decode which terms were sent by an iterative procedure.
%We discuss first the two-step version (subsections A and B). In subsection C we discuss extension to multiple steps.
In the constant power allocation case set $P_j = P/L$.  For the variable power case let $P_j = P_{(\ell)}$ for $j$ in section $\ell$. %We note that each coordinate of $Y$ has expected square $\sigma_Y^2= P+\sigma^2$ and hence $\E[|Y|^2] = P\!+\!\sigma^2$.

\vspace{0.1cm}
\noindent
{\bf First Step:} For each term $X_j$ of the dictionary compute
the statistic $\Zcal_{1,j} = X_j^T Y \,/\|Y\|$.%the inner product with the received string $X_j \cdot Y$ as a test statistic and see if it exceeds a threshold. For our two-step algorithm, the threshold is taken to be a value typically slightly greater than $\sqrt{P/L}$.  Indeed, let $t = \sqrt{P/L} \, (1+a \epsilon)$, where $\epsilon = [R/(2R_{th} \log B)]^{1/2}$ and $a$ is a positive constant.  Later with variable component powers a somewhat different threshold is used. We use a data dependent threshold $T = [|Y|/\sigma_Y] \, t$
%close to $t$.
The terms for which the statistic is above a threshold $\tau = \sqrt{2\log(B)} + \acal$ are regarded as decoded terms. Denote the associated event
$\Hcal_j = \{\Zcal_{1,j} \ge \tau\}$.
The idea of the first step threshold is that very few of the terms not sent will be above threshold. Yet a positive fraction of the terms sent will be above threshold and hence will be correctly decoded on this first step,  %In the constant power allocation case with $R=R_{0}$, this fraction is near $\bar \Phi(\acal) \!=\! 1\!-\!\Phi(\acal)$, the probability that a standard normal is above $\acal$. In the variable power case, the chances of the true terms being above threshold vary for $\ell$ from $1$ to $L$,
with an average likely to be at least a positive value $q$ as will be quantified.

Let $dec_1 =\{j: 1_{\Hcal_j} =1\}$ be the set of terms decoded on this step.
% and $J_1=\{j:1_{\Hcal_j}=0\}$ be the set remaining.  %[In the extremely unlikely event that $DEC_1$ is already at least $L$ there will be no need for the second step.]
The first step provides the fit
$F_1= \sum_j \sqrt{P_j} \, X_j 1_{\Hcal_j}$.  %the course fit
%$Fit \!=\! (P/L)^{1/2}\,\sum_j X_j 1_{\Hcal_j}$.
%Later, for versions with variable component power, we will set $Fit = F = \sum_j X_j \sqrt {P_j} \, 1_{\Hcal_j}$. %with somewhat different threshold in the formation of the events $H_j$.
%[Optionally, for a variant of the algorithm, save the inner products %$X_j \cdot Y$ for subsequent use.]
%and compute $F\cdot Y =(P/L)^{1/2}\, \sum_j X_j\cdot Y 1_{H_j}$.
%Moreover, in preparation for the second step, compute the vector of residuals $$r\,=\, Y\,-\,Fit.$$

\vspace{0.1cm}
\noindent
{\bf Second Step:} %For each of the remaining terms, compute  $\Zcal_{2,j}=X_j^T r \, /\|r\|$.  In the analysis we see, for large $L$ and $B$, it is likely that most of the remaining terms to decode are those for which this newly formed statistic is above a threshold.
%One may see if (the magnitude of) it exceeds a threshold %$\tilde T$ to be specified, %$= \sqrt{P/L}\,(1-b)$, slightly %smaller than $\sqrt{P/L}$ and
%, with a value of the constant $b$ to be specified.
%and accept as decoded those additional terms for which this newly formed statistic is above the smaller threshold.  %Alternatively
%To enforce selection of the specified number of terms on this second step, %one may
%select those for which $\Zcal_{2,j}$ are among the $dec_2$ largest (in magnitude) where $dec_2 = L- dec_1$.  So in fact, for the algorithm, there is no need for a threshold on this second step.
%A variant of this simple second step is now described. It requires slightly more computation, though it illuminates the thinking and further simplifies the resulting analysis.
For each of the remaining terms, form the inner product with the vector of residuals $r=Y-F_1$, that is, compute $X_j^T r$ or its normalized form $\Zcal_j^r=X_j^T r \, /\|r\|$. A quantity with similar properties is found to be equally easy to compute and somewhat simpler to analyze. Indeed, compute  $F_Y=[F_1^T Y/\|Y\|^2]\, Y$ which is the part of $F_1$ in the direction $Y$ and the vector $G= F_1- F_Y$ %$[F\cdot Y/|Y|^2] Y$
which is the part orthogonal to $Y$. For each of the remaining $j$ compute %the inner products $X_j \cdot G$, in normalized form
$\Zcal_{2,j} = X_j^T G \,/\|G\|$.
%, rather than computing the inner products with the residuals.  %Bypass computation of the residual $r$ which would equal $[1-(F\cdot Y)/|Y|^2] Y - G$ and bypass computation of the normalized inner product $\sqrt n X_j \cdot r/|r|$ which would take the form $\sqrt{1-\hat b} \zeta_j + \sqrt{\hat b} \tilde \zeta_j$ with a particular data dependent $\hat b$ given by $|G|^2/|r|^2$.
%Instead, simplification of the analysis and strengthening of the conclusions is obtained by replacing $\hat b$ by the deterministic value $b= q P/(\sigma^2+P)$ and
Then form the combined test statistic
$$\Zcal_{2,j}^{comb} \, = \, \sqrt{1\!-\!\lambda} \, \Zcal_{1,j} \, - \, \sqrt \lambda \, \Zcal_{2,j},$$
with $\lambda= q \, P/(\sigma^2\!+\!P)$.
%, choosing a positive $q\!=\!q_o$ less than $\bar \Phi(\acal)$.
The specified $\lambda$ is chosen to maximize the mean separation between correct and wrong terms. For the two-step version, complete the decoding, in each section not previously decoded, by picking the term for which this statistic is largest, with no need for a second step threshold in that case.
%So in fact, for the algorithm, there is no need for a threshold on this second step.

\vspace{0.1cm}
\noindent
{\bf Extension to Multiple Steps:} We briefly describe how the algorithm is extended to multiple steps
to provide increased reliability. The process initializes with
%the index set $J_0=J$ consisting of all the terms.  Also
$V_{1,j} = X_j$ the vectors of terms in the dictionary with index set $J_1$ consisting of all the terms.  From the first step, $G_1=Y$ is the received vector and the statistics $\Zcal_{1,j}$ are $X_j^T G_1/\|G_1\|$ for $j$ in $J_1$ with associated events $\Hcal_{1,j}=\Hcal_j$. %$\{\Zcal_{1,j} \ge \tau\}$.
%The threshold $\tau_1$ is set to be $\sqrt{2\log B} +\acal$, with $\acal_1$ chosen to keep small the level of false alarms, while allowing some terms to be decoded on the first step.
%and it is also expressed as $\tau_1 = \sqrt{2(R_{th}/R)\log B} + a_1$ with $q_1^* = \bar \Phi(a_1)$.
%As before, the output of the first step consists of the set of decoded terms $dec_1= \{j : 1_{\Hcal_{1,j}} = 1\}$ and the vector $F_1 = \sum_j \sqrt{P_j} \, X_j 1_{\Hcal_{1,j}}$ which forms the first part of the fit.
%and the set $J_1 = J_0 \intersect \{j  :  1_{H_{1,j}^c} =1\}$ of remaining terms.

For the second step the vector $G_2=G$ is formed, which is the part of $F_1$ orthogonal to $G_1=Y$. %, two means of which have been given for its computation.
The set of terms investigated on this step is $J_2 = J_1\intersect \{j  :  1_{\Hcal_{1,j}} =0\}$.
For $j$ in $J_2$, the statistic $\Zcal_{2,j}= X_j^T G_2/\|G_2\|$ is computed as well as the combined statistic $\Zcal_{2,j}^{comb} = \sqrt {\lambda_1} \Zcal_{1,j} - \sqrt{\lambda_2} \Zcal_{2,j}$, where $\lambda_1=1-\lambda$ and $\lambda_2=\lambda$.  What is different on the second step is consideration of the events $\Hcal_{2,j} =\{\Zcal_{2,j}^{comb} \ge \tau\}$ with the same threshold $\tau$, %$\sqrt{2(R_{th}/R)\log B} +a_2$, also expressed as
%$\sqrt{2\log B} + \acal_2$,
leading to an additional part $F_2 = \sum_{j \in J_2} \sqrt{P_j}\, X_j 1_{\Hcal_{2,j}}$ of the fit %$fit_2=$
$F_1+F_2$.
The second step provides some increase in separation, without attempting to resolve all in two steps.
%but rather to decode much of of what remains.

%The terms $X_j$ initially decompose as $X_j = \Zcal_{0,j} G_0/\|G_0\| + V_{1,j}$ with $V_{1,j}= V_{0,j} - \Zcal_{0,j} G_0/\|G_0\|$, previously called $V_j$, with which, for the next step, we form the vector $G_1$, previously called $G$, which is $G_1= \sum_{j\in J_0} V_{1,j} 1_{H_{1,j}}$, based on the remaining set of terms $J_1 = J_0 \intersect \{j : 1_{H_{1,j}^c} = 1\}$, and we form $\Zcal_{1,j}= X_j^T G_1/\|G_1\|$ and the combined statistic $\Zcal_{1,j}^{comb} = \sqrt {\lambda_0} \Zcal_{0,j} - \sqrt{\lambda_1} \Zcal_{1,j}$, where $\lambda_1=\lambda$ and $\lambda_0=1\!-\!\lambda$. Newly consider on this step the events $H_{2,j} =\{\Zcal_{1,j}^{comb} \ge \tau_1\}$ with threshold $\tau_1 \sqrt{2(R_{th}/R)\log B} +a_1$, where $a_1$ is to be specified (perhaps also equal to the same $a$).  The idea is not to attempt to resolve all on the this step, but rather to decode a fraction near $q_1^*=\bar \Phi(a_1)$ of what remains.

Proceed, iteratively, to perform the following loop of calculations, for $k\ge 2$.  From the output of step $k\!-\!1$, there is available the partial fit vector $F_{k-1}$ and for $k' < k$ the previously stored vectors $G_{k'}$ and statistics $\Zcal_{k',j}$ at for $j$ in the previous set $J_{k-1}$.  Plus there is a set $J_k$ of remaining terms for us to consider at step $k$.  %Consider, as discussed further below,
From the residual $r=Y-\hbox{fit}_{k-1}$, one may compute $\Zcal_{k,j}^{res}= X_j^T r/\|r\|$.  Instead, for simplification of the analysis,
compute the part $G_k$ of $F_{k-1}$ orthogonal to the previous $G_{k'}$ and for each $j$ in $J_{k}$ compute $$\Zcal_{k,j}= X_j^T G_k/\|G_k\|$$ and the combined statistic %$$\Zcal_{k,j}^{comb} = \sqrt {\lambda_{1,k}} \,\Zcal_{1,j} - \sqrt{\lambda_{2,k}} \,\Zcal_{2,j} - \ldots - \sqrt{\lambda_{k,k}} \,\Zcal_{k,j},$$ where these $\lambda$'s
$$\Zcal_{k,j}^{comb} = \sqrt{1- \lambda_{k}} \,\Zcal_{k-1,j}^{comb} - \sqrt{\lambda_k} \, \Zcal_{k,j},$$
%where $\lambda_k= w_k/s_k$.
where the value of $\lambda_k$ we shall specify is again chosen to maximize a measure of separation between correct and wrong terms.
 %will be specified with $\sum_{k'=1}^k \lambda_{k',k} = 1$. These positive weights will take the form $\lambda_{k',k} = w_k'/s_k$, with $w_1=1$, and $s_k=1+w_2+\ldots w_k$, with $w_k$ to be specified.  Accordingly, the combined statistic may be computed by the update
%$$\Zcal_{k,j}^{comb} = \sqrt{1- \lambda_{k}} \,\Zcal_{k-1,j}^{comb} - \sqrt{\lambda_k} \, \Zcal_{k,j},$$
%where $\lambda_k= w_k/s_k$.
The statistics $\Zcal_{k,j}^{res}$ are similar, entailing empirically determined values of $\hat \lambda_k$.
The statistics $\Zcal_{k,j}^{comb}$ are compared to the threshold, leading to the events $\Hcal_{k,j} =\{\Zcal_{k,j}^{comb} \ge \tau\}$.
%with $\tau_{k} =$ %$ \sqrt{2(R_{th}/R)\log B} +a_{k}$, also expressed as
%$\sqrt{2\log B} + \acal$, where $\acal_{k}$ is to be specified, perhaps also equal to the same $\acal= \acal_{1}$.
%Along with appending $\Zcal_{k,j}$ to the collection of stored inner products
The output of step $k$ is the vector $$F_{k} = \sum_{j \in J_{k}} \sqrt{P_j} \, X_j 1_{\Hcal_{k,j}},$$ providing the update $\hbox{fit}_k= \hbox{fit}_{k-1}+F_k$.  Also the vector $G_{k}$ and the statistics $\Zcal_{k,j}$ are appended to what was previously stored, at least for the terms $j$ in $J_{k}$.
This step updates the set of decoded terms
$dec_{k}$ to be $dec_{k-1} \union \{j \in J_{k} : 1_{\Hcal_{k,j}}=1\}$ and updates the set of terms remaining for further consideration
$J_{k+1} = \{j \in J_{k} : 1_{\Hcal_{k,j}} = 0\}$.
This completes the actions of step $k$ of the loop. The idea is that on each step $k$ we decode a substantial part
%have decoded a fraction near $q_{k}^*=\bar \Phi(a_{k})$
of what remains, because of growth of the mean separation between terms sent and the others. %as we shall see.

\section{Reliability}

Let $\hat q_{k},\hat f_k$ be the fraction of correct detections and false alarms at step $k$. Also let
$\hat f_{1,k} = \hat f_1 + \ldots + \hat f_k$ be the total fraction of false alarms after $k$ steps.  For the variable power case let $\pi_j=P_j/P$ and use $\hat q_k = \sum_{j \: sent \cap J_k} \pi_j \, 1_{\Hcal_{k,j}}$ and $\hat f_k = \sum_{j \: not \: sent \cap J_k} \pi_j \, 1_{\Hcal_{k,j}}$, as weighted fractions, relative to the total weight of terms sent.

It is not hard to see that $\hat q_{1,k}= \sum_{j \: sent} \pi_j \, 1_{\Hcal_{k,j}}$ is a lower bound on $\hat q_1 +\! \ldots\! + \hat q_k$ the total weighted fraction of correct detections from steps $1$ to $k$.

%The fact that these $\hat q_{1,k}^{adj}$ satisfy the required properties %provides a lower bound on $\hat q_{1}^{adj} \!+\! \ldots \!+\! \hat q_{k}^{adj}$
%is a consequence of the following lemma.
Let's specify a target false alarm rate $f^*$ that arise in our analysis for each step. For step $k$, for given $\acal  >  0$, set
$$f^*=\frac{1}{(\sqrt{2 \log B} + \acal)\sqrt{2\pi}} \exp\{-\acal \sqrt{2 \log B} -(1/2)\acal^2\}$$
%which is smaller than any polynomial in $1/\log B$,
and likewise set values $f > f^*$.  Recall that the threshold $\tau = \sqrt{2\log B}+ \acal$.  Indeed, it is unlikely that $\hat f_k$ exceeds $f$.

Similarly, using distributional properties of $\hat q_{1,k}$ using the function $g(x)$ discussed below, we specify a value $q_{1,k}$ for which we expect that $\hat q_{1,k}$ is likely to be at least $q_{1,k}$.
Further define, $x_0=0$ and for $k\ge 1$,
$$x_k=q_{1,k}^{adj} = \frac{q_{1,k}}{1+ f_{1,k}/q_{1,k}},$$
where $f_{1,k} = k f$.  These $x_k$ are used in setting the weight $\lambda_k$ and in expressing the mean separation $a_{k,j}$ between terms sent and terms not sent.
Indeed $\lambda_k = w_k(1-x_k \nu)$ with
$$w_k = \frac{1}{1-x_k \nu} - \frac{1}{1-x_{k-1}\nu}$$
measuring the increase in a quantity used in specifying the separation.
For establishing reliability, the critical matter is to demonstrate that $x_k=q_{1,k}^{adj}$ grows to a value near $1$. Define
$$\mu_x(u) = \left( \frac{\sqrt{u}}{\sqrt{1-x\nu}} - 1\right)\sqrt{2\log B} - \acal'.$$
Here $\nu= P/(\sigma^2+P)=1-e^{-2\Capacity}$ and $\acal' = \acal + h$, where $h$ is a small number positive number.

The $\Zcal_{k,j}^{comb}$ are not normally distributed, nevertheless, it is demonstrated by induction that in a set of high probability, they are greater than normal random variables which have mean $0$ for terms not sent and mean $a_{k,j}$ for terms sent.
Across the terms $j$, the joint normal distribution that arises in this construction has a covariance $I- \nu_k \beta \beta^T/P$ where $\nu_k\le \nu= P/(P+\sigma^2)$, for which it is shown that the joint density is not more than a constant $1/(1-\nu)^{1/2}=e^\Capacity$ times the joint density that would arise if they were independent standard normal.

In the constant power case with $R=R_{0}$, let $g(x)=\Phi(\mu_x)$.
where $\mu_x=\mu_x(1)$.
%This function has the role
Then for terms sent $a_{k,j} %=a_k
= - \mu_{x_{k-1}}$ and $q_{1,k}^* %= \bar \Phi(a_k)
= g(x_{k-1})$ at $x_{k-1}=q_{1,k-1}^{adj}$.  If $g(x)$ exceeds $x$, then
%at $x=x_{k-1}$ the resulting $q_{1,k}^*$ is greater, so
there is room to set $q_{1,k}$ just below $q_{1,k}^*$, so that if $f_{1,k} = kf$ is small enough, then $x_k=q_{1,k}^{adj}$ is indeed larger than $x_{k-1}$.
%With sufficiently small positive $\eta$, we may set $q_{1,k}=q_{1,k}^*-\eta$. %$q_{1,k-1}^{adj}$.
%Having $R$ less than $R_{th}$ may be used to increase the value of $\mu(x)$ and hence increase whatever gap $g(x)$ has above $x$.  Let's examine here the more stringent case of $R=R_{th}$.  Then
%$$\mu(x)= \sqrt{1/(1-x \nu)}[\sqrt{2\log B} - h] - \sqrt{2\log B} - \acal.$$

%The relevant values of $x$ are $0 \le x \le 1$. In this interval we see that $\mu_x$ and hence $g(x)$ are increasing functions of $x$. Consider the behavior at the end points, $x=0$ and $x=1$.  At $x=0$ we have $\mu_x$ equal to $-\acal'$. Accordingly, at $x=0$, there is a moderate positive value of $g(x)$ equaling $\Phi(-\acal')$. As for the behavior near $x=1$, pick a target value $\gamma$, such as $0.7$ or $0.8$, less than $1$, for which we desire $g(x)$ to have leveled off to be near $1$ for all $x \ge \gamma$.  For such $x\ge \gamma$, the value of $\mu_x$ is at least $[\sqrt{1/(1- \gamma \nu)}-1]\sqrt{2\log B} - \acal'$.  Though $\mu_x$ is only of the order $\sqrt{\log B}$, it produces a value of $g(x)$ that is close to one for such $x$, with normal tail probability not more than $r^*=1\!-\!g(\gamma)$ polynomially small in $1/B$. This $\mu_\gamma$ does not need to be very large to be useful; familiar moderate values such as $1.65$ and $1.96$, respectively, correspond to $r^*=0.05$ and $r^*=0.025$. % lead to $g(\gamma)$ equal to $0.977$ and $0.9985$. As long as $\gamma \nu$ is not too small, such values are reached with suitable $B$.

%Conditions are given such that
The $g(x)-x$ stays above a positive $gap$ %$=g(x^*)-x^*$
for all $0\!\le\! x\! \le\! x^*$.  For the constant power case the positivity holds at $x^*$ provided $x^*$ is separated from $1$ by at least a polynomial in $1/B$, and this gap at $x^*$ is the minimum value in $[0,x^*]$ provided $\acal' \le \sqrt{2\pi}(.5 - \bar x^*)$ and  $\Phi(-\acal') \ge \bar x^*$ where $\bar x^* = 1\!-\!x^*$.

\vspace{0.1cm}
\noindent
{\bf {Lemma 1:}} If $g(x)-x$ is at least a positive $gap$ on an interval $[0,x^*]$, choose small positive $\eta$ and $f > f^*$. Arrange $\Lambda = gap-\eta$ to be positive and for $4f\,x^* \le \Lambda^2$ and arrange $q_{1,k} = q_{1,k}^* - \eta$ where $q_{1,k}^*=g(q_{1,k-1}^{adj})$.  Then the increase on each step $q_{1,k} - q_{1,k-1}$ for which $q_{1,k-1}^{adj} \le x^*$ is at least $\tilde \Lambda$, where $\tL$ satisfies %the equation
$\tL = \Lambda - x^*\,f/\tL$, quadratic in $\tL$ with solution $\tL = \Lambda\{1+ (1\!-\!4\,x^*\,f/\Lambda^2)^{1/2}\}/2$. % near $\Lambda - x_r\,f/\Lambda$.  A slightly larger $\tilde \Lambda$ solving $\tL = \Lambda - x_r\,f/(f+\tL)$ also satisfies $q_{1,k} - q_{1,k-1} \ge \tilde \Lambda$.
Moreover, the number of steps $m$ required such that on step $m-1$, the $q_{1,m-1}^{adj}$ first exceeds $x^*$, is bounded by $m\le 1/\tL$ steps.  At the final step $q_{1,m}$ exceeds $g(x^*)-\eta$.

%Armed with this, arranging suitably small $f$, and $\eta$ %and r^*$ compared to $\bar x^*$, the increase $x_{k}-x_{k-1}$ is assured to be at least a value $\Lambda$ solving $\Lambda = gap -\eta  - f/\Lambda$ for $k$ up to the penultimate step $m\!-\!1$ for which $x_{m-1}$ is at least $x^*$, and hence the $g(x_{m-1})$ used in producing the final $x_m$ is very close to $1$, within a polynomial in $1/B$.  The number of steps required to achieve the indicated performance is not more than $1/\Lambda$.

We also consider the variable power case.  A quantity needed in our analysis is
$C_{\ell,R} = \pi_{(\ell)} \,L\, \nu/2R$.
With $\pi_{(\ell)}$ proportional to $u_{\ell} = e^{-2\Capacity(\ell-1)/L}$, this $C_{\ell,R}$ becomes $u_{\ell} \,\Capacity_L/R$, where the value $\Capacity_L = (L/2)(1-e^{-2\Capacity/L})$ is near the capacity $\Capacity$.  Then $C_{\ell,R}$ is near $u_{\ell}$ when $R$ is near the capacity $\Capacity$.
%which is to be compared to $$u_\ell = e^{-2 \Capacity \frac{\ell-1}{L}}$$ where $\pi_{(\ell)} = P_{(\ell)}/P$. In determining the rate condition such that the fraction correctly decoded becomes close to $1$, it is found that the critical matter is, for all $\ell$ from $1$ to $L$, to have $C_{\ell,R}$ at least $u_\ell$, or slightly higher in the variable case.  In the constant power case $\pi_{(\ell)} = 1/L$, this condition means $R_{th}/R \ge 1$.  For the variable power case, which requires an additional refined argument, our condition is
%$$C_{\ell,R} \, \ge \, \big(\sqrt{u_\ell}+\epsilon_B)^2$$ where $\epsilon_B = \acal/\sqrt{2\log B}$.  For instance with $\pi_{(\ell)}$ proportional to the right side the largest rate for which the condition is satisfied is essentially $$R \:= \:  \frac{\Capacity}{1+ 4 \,\epsilon_B \,\rho \, + 2 \,\epsilon_B^2 \, \Capacity/\nu}.$$
In the variable power case, the mean separation of the $\Zcal_{k,j}$ is given by $a_{k,j_\ell}=-\mu_x(C_{\ell,R})$ for section $\ell$.  Likewise the role of the function $g(x)$ is played by
$$g_L(x)=\sum_{\ell=1}^{L} \pi_{(\ell)} \, \Phi(\mu_x(C_{\ell,R})).$$
When $\pi_{(\ell)}$ is proportional to $u_\ell=e^{-2\Capacity (\ell-1)/L}$ this $g_L(x)$ is at least the value of a nearby integral
%has a value nearly equal to the integral
$$g(x)= \frac{1}{\nu} \int_{e^{-2C}}^1 \Phi(\mu_x(u \Capacity/R))\,du.$$
%with $\mu_x^*(u) = \big(\sqrt {u/(1-x\nu)} - 1\big)\sqrt{2\log B}$.
%Under restrictions on the size permitted for $\acal'$,
This $g(x)$ is found to compare favorably to $x$, to yield the required growth of the $x_k$. %, with a number of steps of order $\log B$. %$\sqrt{\log B}/\acal'$.  Furthermore, to allow $\acal'=\acal+h$ formed from any positive $\acal$ and $h$, an additional argument is developed, involving contributions to correct decoding on each step $k$ from wedges formed in the two-dimensions $\Zcal_{k-1,j}^{comb},\Zcal_{k,j}$ by the intersection of half-spaces $\Hcal_{k,j}^c \intersect \Hcal_{k-1,j}$.

Consider the case allowing leveling with which $\pi_{(\ell)}= \max\{u_\ell,cut\}/ sum$, for which the normalizing sum is found to be $(L\nu/2\Capacity) [1+\delta_{sum}^2]$, where $\delta_{sum}^2$ is near $D(\delta_c)/snr$, bounded by $\delta_c^2/(2snr)$, with $\delta_c=c/\sqrt{2\log B}$ and $D(\delta)=(1\+\delta)\log(1\+\delta)-\delta$.  The function $g_L(x)$ is defined as above with an analogous nearby integral with $\max\{u,cut\}$ in place of $u$.
%with allowance for leveling is defined in a manner analogous to the above.
Set $r>0$ and consider the rate
$$R= \frac{\Capacity}{(1+\delta_{sum}^2)(1+\delta_a)^2(1+2r/\tau_B^2)},$$
where $\tau_B^2= 2(\log B)(1 \+ \delta_a)^2$ with $\delta_a = a'/\sqrt{2\log B}$.  %Set the value of $a$ to achieve a suitably small false alarm rate $f^*$ so that $f=2f^*$ equals $gap^2/4$.
Setting a suitably small false alarm rate to not interfere with the accumulation of correct detections, the resulting $\delta_a$ is of order
$[\log \log B + \log snr ]/(\log B)$, so all three sources of rate drop above, $\delta_{sum}^2$, $\delta_a$ and $r/\tau_B^2$ are of order $1/\log B$ to within a loglog factor.  A relevant lemma is the following.

\vspace{0.1cm}
\noindent
{\bf Lemma 2:}  Let $x_{up}$ be near $1$ with $1-x_{up} = (1/snr)(2r/\tau_B^2)$. For any non-negative $a$, $c$, and $r$, with the rate given above, the function $g(x)-x$ for $0 \le x \le x_{up}$, is minimized at $x_{up}$.

\vspace{0.1cm}
The proof is based on an evaluation of the integral $g(x)$ which has expression in terms of the variable $z=\mu_x(cut \,\Capacity/R)$ which is one-to-one with $x$.  The value $x_{up}$ corresponds to a point $z_{up}=\zeta$ with favorable properties. %where $1\+\delta_c = (1\+\zeta/\tau)^2$, with $\zeta=0$ when $c=0$.
Expressing the function in terms of $z$, one makes separate treatment of the behavior for $z\le -\zeta$, where the function is close to decreasing, and for $-\zeta \le z \le +\zeta$, where the function is close to symmetric, slightly skewed to be lower at $+\zeta$.

The value of $\zeta$ is near $c/2$. Consider choices that approximately optimize the overall rate, yielding $\zeta$ near $\sqrt{\log_+((\log B)/4\pi)}$, at which the $gap$ of $g(x)-x$ at $x_{up}$ is at least a value near $(1/snr) (2r-1/2)/\tau_B^2$, positive for $r > 1/4$. Moreover, choosing $a$ such that the false alarm rate $f=2f^*$ equals $(gap-\eta)^2/4$, so that the conditions of Lemma 1 are satisfied, it produces a value of $\delta_a$ of the indicated form. %of the indicated order.

%We are now in a position to
Let's state the result regarding reliability of the multi-step adaptive successive decoder.  The proof is based on the above-mentioned normal approximation bound and a large deviation bound for weighted combinations of Bernoulli random variables, for which one may see the full manuscript \cite{BarronJosephFast}.  %For specificity of the error probability bounds, it is stated below in the equal power case. Analogous exponential bounds hold for the variable power case under the above condition on $C_{\ell,R}$, which allows rates $R$ up to the value near capacity $\Capacity$ as indicated.

\vspace{0.1cm}
\noindent{\bf Theorem 3:}
Suppose the communication rate and power allocation are such that $g$ is accumulative, with $g(x)-x > 0$ on $[0, x^*]$. Pick $\eta_k = \eta$ and $f \!>\! f^*$ such that the conditions of Lemma 1 are satisfied, or more generally arrange
$q_{1,k} = g(q_{1,k-1}^{adj})-\eta_{k}$ so
%For a false alarm rate target $f \!>\! f^*$, and failed detection rate targets  $r^*\!=\!1\!-\!g(x^*)$, set recursively $q_{1,k} = g(q_{1,k-1}^{adj})-\eta_{k}$, with $q_{1,k}^{adj} = q_{1,k}/[1+f_{1,k}/q_{1,k}]$, where $f_{1,k}=kf $ and $\eta_{k}= q_{1,k}^*-q_{1,k}>0$. If the $f_{1,k}$ and $\eta_k$, as well as the quantities that arise in the definition of $g$, are chosen such
that the increase $q_{1,k}-q_{1,k-1}$ remains positive
%bounded away from zero
for $k \!<\! m$.
%if $\acal+h \le \sqrt{2\pi}(0.5-\bar \gamma)$ and $\Phi(-(\acal+h)) \ge \bar \gamma$ and if $f_{1,k'} = \sum_{k''=1}^{k'} f_{k''}$ and $\tau_k'= q_{1,k'}^*-q_{1,k'}$ are such that $\bar \gamma - r^* - \tau_k' - f_{1,k'}$, which lower-bounds the increase $q_{1,k'}-q_{1,k'-1}$, remains bounded away from zero, where $q_{1,k'} = g(q_{1,k'-1}^{adj})-\tau_k'$,
If the penultimate step $m\!-\!1$ is such that $q_{1,m-1}^{adj}$ is the first with value at least $x^*$, then with $rem=1-q_{1,m}$, the $m$ step single-dictionary decoder incurs a fraction of errors less than $mf + rem$, except in an event of probability not more than the sum for $k$ from $1$ to $m$ of
$$e^{-L_\pi D(q_{1,k}\|q_{1,k}^*)+c_0 k} + e^{-L_\pi (B\!-\!1)D(p\|p^*)} + e^{-(n\!-\!k\!+\!1)D_{\epsilon_k}}.$$
%+ \sum_{k'=1}^k e^{-(n\!-\!k'\!+\!1)D(\epsilon_{k'})}.$$
%\PP\big[\Chi_{n-1}^2 \le 1 - h/\sqrt{C_{R,B}}\,\big],$$
%where the terms correspond, respectively, to the fractions of detection on steps $1$ to $k$, %$k\!-\!1$,
%the fractions of false alarms on steps $1$ to $k$, %the final fraction of failed detections,
%and the tail probabilities for $\|G\|_{k'}^2/n \le (1- h/\sqrt{C_{R,B}})^2$.
Here  $D(\cdot\|\cdot)$ refers to the Kullback-Leibler divergence between two Bernoulli random variables; $p,p^*$ equal the corresponding $f,f^*$ divided by $B\!-\!1$; and  $D_\epsilon= -\log(1-\epsilon) - \epsilon$ which is at least
$\epsilon^2/2$. Also $\epsilon_{k} = (n \epsilon \!-\!k\!+\!1)/(n\!-\!k\!+\!1)$, where $\epsilon=1-(1-h/\sqrt{2\log(B)}\,)^2$, and $c_0=\Capacity$. Moreover, $L_{\pi} = 1/\max_{\ell} \pi_{\ell}$, approximately a constant multiple of $L$ for the designs investigated here.
%With $\eta$ constant and $f=cf^*$ proportional to $f^*$ with $c > 1$, the following simplified bound also holds $$k \, e^{-2L\eta^2+c_0 k} + k \, e^{-Lf^* D_c} + k \, e^{-(n\!-\!k\!+\!1)D(\epsilon_{k})}.$$ where $D_c =c\log c -(c-1)$ is positive.
%where $D_f = f \log f/f^* - (f-f^*)$.
%This bound is nearly exponentially small in $L$, since $f^*$ goes to zero slower than any polynomial in $1/B$.

To produce each step $q_{1,k}$ from $q_{1,k}^*$, one may set a constant difference $\eta_k\!=\!\eta$ and invoke the bound $D(q_{1,k}\|q_{1,k}^*)\ge 2\eta^2$.
A preferred tactic, used in producing the curves shown earlier, is each step to choose  $q_{1,k}$ to produce constancy of the exponent $D(q_{1,k}\|q_{1,k}^*)$ at a prescribed value, equalizing the contributions to the above probability bound from each step.

\section*{Acknowledgment}
% optional entry into table of contents (if used)
%\addcontentsline{toc}{section}{Acknowledgment}

%We thank Dan Spielman, Edmund Yeh, John Hartigan, Mokshay Madiman, Imre Teletar, Cong Huang, Xi Luo, Harrison Zhou, Joseph Chang, for helpful conversations, along with David Smalling and
Creighton Heaukulani is thanked for helpful simulations.
%for projects in his masters program in Statistics at Yale.
%who were especially helpful in providing simulation of earlier forms of the decoders.
%We especially thank David Smalling who completed a number of simulations of earlier incarnations of the decoding algorithm for his Yale applied math senior project \cite{Smalling2009} in spring term of 2009 and Yale statistics masters student Creighton H. who took the simulations further in the summer of 2009, their work was instrumental to us in recognizing that without modification direct convex projection methods have a rate threshold substantially below capacity when the signal-to-noise ratio is high.

%\bibliographystyle{/Users/mokshay/Documents-ACADEMIC/WRITINGS/CommonResources/IEEEtranBST/IEEEtran}
%\bibliography{/Users/mokshay/Documents-ACADEMIC/WRITINGS/CommonResources/poi,/Users/mokshay/Documents-ACADEMIC/WRITINGS/CommonResources/ik}

\end{document}